\documentclass[copyright]{eptcs}
\providecommand{\event}{T. Neary, D. Woods, A.K. Seda and N. Murphy (Eds.):\\ The Complexity of Simple Programs 2008.}
\providecommand{\volume}{1}
\providecommand{\anno}{2009}
\providecommand{\firstpage}{172}
\usepackage{breakurl}

\usepackage{graphicx}

\usepackage{xspace}
\usepackage[basic,classfont=bold]{complexity}
\usepackage{amssymb}
\usepackage{amsmath}
\usepackage{amsthm}

\usepackage{amsfonts}
\usepackage{amssymb}
\usepackage{bm}

\newtheorem{thm}{Theorem}
\newtheorem{lem}[thm]{Lemma}

\newtheorem{defn}[thm]{Definition}
\newtheorem{rem}[thm]{Remark}
\newtheorem{prob}[thm]{Problem}

\newlang{\STCON}{STCON}
\newlang{\NOTSTCON}{\overline{\STCON}}
\newlang{\GAP}{GAP}
\newlang{\DFA}{DFA}
\newlang{\RUSTCON}{RUSTCON}
\newlang{\ACCESSIBLE}{ACCESSIBLE}
\newclass{\DLOGTIME}{DLOGTIME}
\newclass{\RUL}{RUL}
\newlang{\GENERALREACH}{GENREC}
\newlang{\STANDARDREACH}{STDREC}
\newlang{\RESTRICTEDREACH}{RSTREC}

\newcommand{\inputnode}{{\tt in}\xspace}
\newcommand{\yes}{{\tt yes}\xspace}
\newcommand{\no}{{\tt no}\xspace}
\newcommand{\DG}{\ensuremath{\mathcal{G}}}
\newcommand{\DGv}{\ensuremath{V_{\DG}}}
\newcommand{\DGe}{\ensuremath{E_{\DG}}}

\newcommand{\nmMembraneObjects}{\ensuremath{O}\xspace}           
\newcommand{\nmGenericObject}{\ensuremath{o}\xspace}             

\newcommand{\PMCuni}          {\ensuremath{\mathbf{PMC}}\xspace}
\newcommand{\PMCsemi}         {\ensuremath{\mathbf{PMC}^{*}}\xspace}

\newcommand{\ACPMCwodissSemi} {\ensuremath{(\AC^0)\text{--}\PMCsemi_{\mathcal{AM}^{0}_{-d}}}\xspace}

\title{On acceptance conditions for membrane systems: characterisations of {L} and {NL}}

\author{Niall Murphy\footnote{Funded by the Irish Research Council for Science, Engineering and Technology}
    \email{nmurphy@cs.nuim.ie}
\institute{Department of Computer Science, National University of Ireland Maynooth, Ireland }
\and
Damien Woods\footnote{Funded by Junta de Andaluc{\'i}a grant TIC-581}
\institute{Deptartment of Computer Science \& Artificial Intelligence, University of Seville,~Spain}
\email{dwoods@us.es}
}
\def\titlerunning{On acceptance conditions for membrane systems: characterisations of {L} and {NL}}
\def\authorrunning{N. Murphy and D. Woods}

\begin{document}
    \maketitle

    \begin{abstract}
        In this paper we investigate the affect of various acceptance conditions on recogniser membrane systems without dissolution.
        We demonstrate that two particular acceptance conditions (one easier to program, the other easier to prove correctness) both characterise the same complexity class, $\NL$.
        We also find that by restricting the acceptance conditions we obtain a characterisation of $\L$.
        We obtain these results by investigating the connectivity properties of dependency graphs that model membrane system computations.
    \end{abstract}


    \section{Introduction}
        In the membrane systems (also known as P-systems~\cite{Pau2002x}) computational complexity community it is common practice to explore the power of systems by allowing and prohibiting different developmental rules.
        This technique has yielded several interesting results such as the role of membrane dissolution in recognising $\PSPACE$-complete problems~\cite{NJNC2006p} and the role of membrane division in recognising problems outside of $\P$~\cite{ZFM2000c}.

        In this paper we do not vary the rules permitted in membrane systems but instead we vary the acceptance conditions and observe the change (or lack of change) this makes to the computing power of the system.
        Our main technique is to analyse the structure, and connectivity, of dependency graphs~\cite{NJNC2006p} that are induced by acceptance conditions.
        Our approach builds on previous work on dependency graphs~\cite{NJNC2006p,GPRR2006c} to give a number of new techniques and results.
        Our techniques and results should be of interest to those who wish to characterise complexity classes, those studying acceptance conditions for membrane systems, and those characterising the power of membrane systems.

        This research was motivated by the realisation that in prior work~\cite{MW2008c} we were using a seemingly more general halting condition than is used by the membrane community.
        Previously, we showed that $\AC^0$-uniform families\footnote{ All membrane systems in this paper are $\AC^0$-uniform and run for polynomial time.} of active membrane systems without dissolution, and using the acceptance conditions specified in Section~\ref{ssec:general_rec_sys}, characterise $\NL$~\cite{MW2008c}.
        However, most researchers use a more restricted acceptance condition (see Section~\ref{ssec:standard_rec_sys}).
        We show here that this more restricted definition also characterises $\NL$.
        This means that the two definitions are equivalent in terms of computing power for \ACPMCwodissSemi systems.
        The choice of which definition to use is now mostly a matter of personal taste as we have shown that the two are equivalent under $\AC^0$ reductions, i.e.\ there is a (very efficient) compiler to translate one definition to another.

        In Section~\ref{ssec:strict_rec_sys} we show that active membrane systems without dissolution, and using a restriction on the standard acceptance definition, characterise $\L$.
        This demonstrates that not all (minor) restrictions on halting definitions yield systems that characterise $\NL$.
        
        We note here that the three definitions that we consider in Section~\ref{sec:different_accepting} all characterise $\P$ if they are generalised to use $\P$-uniformity.
        The $\P$ lower bound of this characterisation is a trivial corollary of the fact that such membrane systems can easily embed polynomial time deterministic Turing machines, and is not related to the differences in their definitions.

    \section{Preliminaries}
        \label{sec:definitions}
        In this section we define membrane systems and some complexity classes.
        These definitions are based on those from P{\u a}un~\cite{Pau2002x,Pau2001p}, Sos{\' i}k and Rodr{\'i}guez-Pat{\'o}n~\cite{SR2007p}, Guti{\'e}rrez-Naranjo et al.~\cite{NJNC2006p}, and  P{\'e}rez-Jim{\'e}nez et al.~\cite{PRS2003p}.
        Previous works on complexity and membrane systems spoke of solving a problem in a ``uniform way'', that is, in a manner reminiscent of how families of circuits solve a problem. 
        Sos{\' i}k and Rodr{\'i}guez-Pat{\'o}n defined uniformity for membrane systems in a similar manner to circuit uniformity, this allows us to refer to uniform families of membrane systems.

        \subsection{Active membrane systems}
            Active membrane systems are a class of membrane systems with membrane division rules.
            Division rules can either only act on elementary membranes, or else on both elementary and non-elementary membranes.
            An elementary membrane is one which does not contain other membranes (a leaf node, in tree terminology).

            \begin{defn}
                \label{def:membrane}
                An active membrane system without charges is a tuple $\Pi = (\nmMembraneObjects, H, \mu, w_1, \ldots, w_m, R)$ where,
                \begin{enumerate}
                        \item $m \geq 1$ is the initial number of membranes;
                        \item \nmMembraneObjects is the alphabet of objects;
                        \item $H$ is the finite set of labels for the membranes;
                        \item $\mu$ is a membrane structure in the form of a tree, consisting of $m$ membranes (nodes), labelled with elements of $H$.
                        The parent of all membranes (the root node) is called the ``environment'' and has label $env \in H$;
                        \item $w_1, \ldots, w_m$ are strings over \nmMembraneObjects, describing the multisets of objects placed in the $m$ regions of $\mu$.
                        \item $R$ is a finite set of developmental rules, of the following forms:
                            \begin{enumerate}
                                \item $[\ a\ \rightarrow\ u\ ]_h$, for $h \in H,\ a \in \nmMembraneObjects,\ u \in \nmMembraneObjects^{*}$
                                
                                \item $a[\ ]_h \rightarrow [\ b\ ]_h$, for $h \in H,\ a,b \in \nmMembraneObjects$
                                
                                \item $[\ a\ ]_h \rightarrow [\ ]_h\ b$, for $h \in H,\ a,b \in \nmMembraneObjects$
                                
                                \item $[\ a\ ]_h \rightarrow b$, for $h \in H,\ a,b \in \nmMembraneObjects$

                                \item[($e$)] $[\ a\ ]_h \rightarrow [\ b\ ]_h\ [\ c\ ]_h$, for $h \in H,\ a,b,c \in \nmMembraneObjects$.
                                
                                \item[($f$)] $[\ a\ [\ ]_{h_1}\ [\ ]_{h_2}\ [\ ]_{h_3}\ ]_{h_0} \rightarrow [\ b\ [\ ]_{h_1}\ [\ ]_{h_3} ]_{h_0}\ [\ c \ [\ ]_{h_2}\ [\ ]_{h_3} ]_{h_0}$,\\
                                for $h_0, h_1, h_2, h_3 \in H,\ a,b,c \in \nmMembraneObjects$.
                            \end{enumerate}
                \end{enumerate}
            \end{defn}
            These rules are applied according to the following principles:
            \begin{itemize}
                    \item All the rules are applied in a maximally parallel manner.
                    That is, in one step, one object of a membrane is used by at most one rule (chosen in a non-deterministic way), but any object which can evolve by one rule of any form, must evolve.
                    \item If at the same time a membrane labelled with $h$ is divided by a rule of type ($e$) or ($f$) and there are objects in this membrane which evolve by means of rules of type ($a$), then we suppose that first the evolution rules of type ($a$) are used, and then the division is produced.
                    This process takes only one step.
                    \item The rules associated with membranes labelled with $h$ are used for membranes with that label.
                    At one step, a membrane can be the subject of only one rule of types ($b$)--($f$).

                    \item Rules of type ($f$) are division rules for non-elementary membranes.
                        These rules allow us duplicate an entire branch of the membrane structure in the following manner.
                        If the membrane (label $h_0$) to which the non-elementary division rule is applied contains objects and child membranes then copies of those membranes and all of their contents (including their own child membranes) are found in both resulting copies of $h_0$.
            \end{itemize}

        \subsection{Recogniser membrane systems}
            In this paper one of our goals is to unify and clarify definitions for language recognising variants of membrane systems.
            To achieve this, we consider three different notions of acceptance for recogniser systems, one in each of Sections~\ref{ssec:general_rec_sys} to~\ref{ssec:strict_rec_sys}.
            Each of these three definitions is a restriction on the general (and purposely vague) Definition~\ref{def:vague_recogniser} below.
            
            We recall from~\cite{NJNC2006p} that a computation of the system is a sequence of configurations such that each configuration (except the initial one) is obtained from the previous one by a transition.
            A computation that reaches a configuration where no more rules can be applied to the existing objects and membranes is called a halting computation.
            
            \begin{defn}
                \label{def:vague_recogniser}
                A {\em recognizer membrane system} is a membrane system with external output (that is, the results of halting computations are encoded in the environment) such that:
                \begin{enumerate}
                    \item the working alphabet contains two distinguished elements \yes and \no;
                    \item if $C$ is a computation of the system, then it is either an accepting or a rejecting computation.
                \end{enumerate}
            \end{defn}
            
            This definition is vague since we have not defined accepting and rejecting computations.
            In Section~\ref{sec:different_accepting} we show the set of problems that a membrane system accepts when using various notions of accepting (or rejecting) computations.

        \subsection{Complexity classes}
            Consider a decision problem $X$, i.e.\ a set of instances $X = \left\{x_1,x_2,\ldots \right\}$ over some finite alphabet such that to each~$x_i$ there is an unique answer ``yes'' or ``no''.
            We say that a {\em family} of membrane systems solves a decision problem if each instance of the problem is solved by some family member.
            We denote by $|x| = n$ the length of any instance $x \in X$.
			Throughout this paper, $\AC^0$ circuits are $\DLOGTIME$-uniform, polynomial sized (in input length $n$), constant depth, circuits with AND, OR and NOT gates, and unbounded fanin~\cite{BIS1990p}.
            The complexity class $\L$ ($\NL$) is the set of problems solved by (non-)deterministic Turing machines using only $O(\log n)$ space, where $n$ is the length of the input instance.

            \begin{defn}
                \label{def:uniformFam}
                Let $\mathcal{D}$ be a class of membrane systems and let $f:\mathbb{N} \rightarrow \mathbb{N}$ be a total function. 
                The class of {\em problems solved by $\AC^0$-uniform families of membrane systems} of type $\mathcal{D}$ in time $f$, denoted $(\AC^0)\text{--}{\bf MC}_{\mathcal{D}}(f)$, contains all problems $X$ such that:
                \begin{itemize}
                    \item There exists an $\AC^0$-{\em uniform} family of membrane systems, $\bm{\Pi}_X = (\Pi_X(1),\Pi_X(2),\ldots)$ of type $\mathcal{D}$: that is, there exists an $\AC^0$ circuit family such that on unary input~$1^n$ the $n^{\mathrm{th}}$ member of the circuit family constructs $\Pi_X(n)$. 
                    \item There exists an $\AC^0$ circuit family such that on input $x \in X$, of length $|x|=n$, the $n^{\mathrm{th}}$ member of the family encodes $x$ as a multiset of input objects placed in the distinct input membrane of $\Pi_{X}(n)$.
                    \item $\bm{\Pi}_X$ is {\em sound} and {\em complete} with respect to problem $X$: $\Pi_X(n)$ starting with an encoding of input $x \in X$ of length $n$ accepts iff the answer to $x$ is ``yes''.
                    \item $\bm{\Pi}_X$ is $f$-efficient: $\Pi_X(n)$ always halts in at most $f(n)$ steps.
               \end{itemize}
           \end{defn}
            
            Definition~\ref{def:uniformFam} describes $\AC^{0}$-uniform families and we generalise this to define {\em $\AC^{0}$-semi-uniform families of membrane systems} $\bm{\Pi}_X =$ $(\Pi_X(x_1);$ $\Pi_X(x_2);\ldots)$ where there exists an $\AC^0$ circuit family which, on an input $x \in X$, constructs membrane system $\Pi_X(x)$.
            Here a single circuit family (rather than two) is used to construct the semi-uniform membrane family, and so the problem instance is encoded using objects, membranes, and rules. 
            In this case, for each instance of $x \in X$ we have a special membrane system which does not need a separately constructed input.
            The resulting class of problems is denoted by $(\AC^0)\text{--}\mathbf{MC}^{*}_{\mathcal{D}}(f)$.
            Obviously, $(\AC^0)\text{--}\mathbf{MC}_{\mathcal{D}}(f) \subseteq (\AC^0)\text{--}\mathbf{MC}^*_{\mathcal{D}}(f)$ for any given class $\mathcal{D}$ and a valid~\cite{BDG1988x} complexity function $f$.


            We define $(\AC^0)\text{--}\PMCuni_{\mathcal{D}}$ and $(\AC^0)\text{--}\PMCsemi_{\mathcal{D}}$ as
            \begin{equation*}
                (\AC^0)\text{--}\PMCuni_{\mathcal{D}} = \bigcup\limits_{k\in\mathbb{N}}(\AC^0)\text{--}\mathbf{MC}_{\mathcal{D}}(n^{k}),\ 
            \end{equation*}\\
            and
            \begin{equation*}
                (\AC^0)\text{--}\PMCsemi_{\mathcal{D}} = \bigcup\limits_{k\in\mathbb{N}}(\AC^0)\text{--}\mathbf{MC}^{*}_{\mathcal{D}}(n^{k}).
            \end{equation*}
            In other words, $(\AC^0)\text{--}\PMCuni_{\mathcal{D}}$ (and $(\AC^0)\text{--}\PMCsemi_{\mathcal{D}}$) is the class of problems solvable by uniform (respectively semi-uniform) families of membrane systems in polynomial time.
            We let $\mathcal{AM}^{0}$ denote the class of membrane systems with active membranes and no charges.
            We let $\ACPMCwodissSemi$ denote the class of problems solvable by $\AC^0$-semi-uniform families of membrane systems in polynomial time with no dissolution rules.
            In an abuse of notation, we often let \ACPMCwodissSemi refer to the class of such membrane systems (rather than problems).
            For brevity we often write $\Pi_X$ instead of~$\Pi_{X}(n)$ or $\Pi_{X}(x)$.

            \begin{rem}
                \label{rem:confluent}
                A membrane system is {\em confluent} if it is both sound and complete.
                That is a $\Pi_X$ is {\em confluent} if all computations of $\Pi_X$ with the same input give the same result; either always accepting or else always rejecting. 
            \end{rem}
            In a confluent membrane system, given a fixed initial configuration, the system non-deterministically chooses one from a number of valid configuration sequences, but all of the reachable configuration sequences must lead to the same result, either all accepting or all rejecting.

        \subsection{Dependency graphs and normal forms}
            The {\em dependency graph} (first introduced by Guti{\'e}rrez-Naranjo et al.~\cite{NJNC2006p}) is an indispensable tool for characterising the computational complexity of membrane systems without dissolution.
            This technique is reminiscent of configuration graphs for Turing Machines.
            Similarly to a configuration graph, a dependency graph helps visualise a computation.
            However, it differs in its approach by representing a membrane system configuration as a set of nodes rather than as a single node in configuration space.

            Looking at membrane systems without dissolution as dependency graphs allows us to employ the existing, mature corpse of techniques and complexity results for graph problems.
            As we show in this paper, this greatly simplifies the process of proving upper and lower bounds for such systems.
            A key technique we use in this paper is to transfer from a dependency graph to a new membrane system, $\Pi \rightarrow \DG_{\Pi} \rightarrow \Pi_{\DG_{\Pi}}$. 
            This new system accepts iff the original membrane system accepts, since their dependency graphs are isomorphic.
            Also, the new system is considerably simplified as it uses only one membrane (the environment) and all rules are of type~($a$).
            This is used as a normal form for membrane systems without dissolution.
            
            In Sections~\ref{ssec:general_rec_sys} to \ref{ssec:strict_rec_sys} we define reachability problems for dependency graphs such that if the answer to the graph reachability problem is yes, then the membrane system it represents is an accepting system.
            This is because the nodes of a dependency graph represent an object being in a certain membrane, and an edge between two nodes represents a developmental rule that causes that object to be in that membrane.
            Thus if the object $\yes$ arrives in the environment (the acceptance signal) of the membrane system, then there is a directed path leading from one special node ($\inputnode$) to another special node ($\yes$) in the dependency graph.
            For more details about how a dependency graph is constructed and its proof of correctness see Guti{\'e}rrez-Naranjo et al.~\cite{NJNC2006p,GPRR2006c}.
            
            The dependency graph for a membrane system $\Pi$ is a directed graph $\DG = (\DGv, \DGe, \inputnode, \yes, \no)$ where $\inputnode \subseteq \DGv$ represents the input multiset, and $\yes, \no \in \DGv$, represent the accepting and rejecting signals respectively.
            Each vertex $a \in \DGv$ is a pair $a=(\nmGenericObject, h) \in \nmMembraneObjects \times H$, where $\nmMembraneObjects$ is the set of objects in $\Pi$ and $H$ is the set of membrane labels in $\Pi$.
            An edge $(a,b)$ exists iff there is a developmental rule in $\Pi$ such that the left hand side of the rule has the same object-membrane pair as $a$ and the right hand side has an object-membrane pair matching $b$.
            Since there is no membrane dissolution allowed, the parent/child relationships of membranes does not change during the computation.
            This allows us to determine the correct parent and child membranes for type ($b$) and type ($c$) rules.

            Previously~\cite{NJNC2006p}, the graph $\DG$ was constructed from~$\Pi$ in polynomial time.
            We make the observation that the graph $\DG$ can be constructed in $\AC^0$.
            We use a common circuit technique known as ``masking'' whereby using AND gates and a desired pattern we filter out the bits of the input string that we are interested in.
            We take as input a binary string $x$ that encodes a membrane system, $\Pi$.
            To make a dependency graph from a membrane system requires a constant number of parallel steps that are as follows. 
            First, a row of circuits identifies all type ($b$) and ($c$) rules and uses the membrane structure to determine the correct parent membranes, then writes out (a binary encoding of) edges representing these rules.
            Next, a row of circuits writes out all edges representing type ($e$) and ($f$) rules (see~\cite{NJNC2006p} for more details about the representation of these rules in dependency graphs).
            For ($a$) rules it is possible to have polynomially many copies of polynomially many distinct objects on the right hand side of a rule.
            To write out edges for these rules in constant time we take advantage of the fact that we require at most one edge for each object-membrane pair in $\nmMembraneObjects \times H$. 
            We have a circuit for each element of $\{ o_h \mid o \in O, h \in H \}$.
            The circuit for $o_h$ takes as input (an encoding of) all rules in $R$ whose left hand side is of the form $ [ o ]_h $.
            The circuit then, in a parallel manner, masks (an encoding of) the right hand side of the rule (for example $ [ bbcdc ]_{h}$) with the encoding of each object in $\nmMembraneObjects$, (in the example, masking for (encoded) $b$ would produce (encoded) $bb000$).
            All encoded objects in the string are then ORed together so that if there was at least one copy of that object in the system we obtain a single instance of it.
            The circuit being unique for a specific left hand side $[ o ]_h$ now writes out an encoding of the edge $(o_h, b_h)$ and an encoding of all other edges for objects that existed on the right hand side of this rule in parallel.

            \begin{rem}
                \label{rem:dg_to_memsys}
                Of course one can take the opposite view.
                We observe that to convert a dependency graph $\DG=(\DGv,\DGe, \inputnode, \yes, \no)$ into a new membrane system, $\Pi_\DG$, we simply convert the edges of the graph into object evolution rules.
                The set of objects of $\Pi_\DG$ is $O_\DG = \DGv$.
                The rules of $\Pi_\DG$ are\linebreak $\left\{ \left[\,v \rightarrow S(v) \right]_{env} \vert\ \forall\ v \in \DGv\right\}$ where $ S(v) = \left\{ s \in \DGv \vert (v,s) \in \DGe\right\}$.
                The nodes $\inputnode, \yes, \no$ become the input multiset, $\yes$ object, and $\no$ object respectively.
                We compute this in $\AC^0$.
            \end{rem}

            This new membrane system, $\Pi_\DG$, highlights some points about active membrane systems without dissolution.
            These give rise to significant simplifications and normal forms.

            \begin{lem}
                \label{lem:pmcwodiss_one_mem}
                Any $\ACPMCwodissSemi$, $\Pi$, with $m$ membranes can be simulated by a \ACPMCwodissSemi system, $\Pi'$, that (1) has no membranes other than the environment and (2) uses only rules of type ($a$).
            \end{lem}
            By {\em simulate} we mean that the latter system accepts on input $\inputnode$ iff the former does. 
            To see that Lemma~\ref{lem:pmcwodiss_one_mem} holds, first notice how the dependency graph represents an (object, label) pair as a single node.
            Also if we convert the dependency graph $\DG$ into a membrane system $\Pi_{\DG}$, (1) it uses a single membrane with label {\em env}, and each node is modelled by a single object.
            (2) Each edge in $\DG$ becomes a rule of type~($a$).
            Notice that the dependency graphs of $\Pi$ and $\Pi_{\DG}$ are isomorphic.

            \begin{lem}
                \label{lem:pmcwodiss_multi}
                Any \ACPMCwodissSemi system, $\Pi$, which has, as usual, multisets of objects in each membrane can be simulated by another \ACPMCwodissSemi system, $\Pi'$, which has sets of objects in each membrane.
            \end{lem}
            
            We verify Lemma~\ref{lem:pmcwodiss_multi} by observing that in a dependency graph, $\DG$, the multiset of objects is encoded as a {\em set} of vertices, no information is kept regarding object multiplicities.
            Thus when $\DG$ is converted into a new membrane system, $\Pi_{\DG}$, there are no rules with a right hand side with more than one instance of each object.
            The resulting system $\Pi_\DG$ accepts iff $\Pi$ accepts since the dependency graphs of both systems are isomorphic. 
            Thus object multiplicities do not affect whether the system accepts or rejects.

    \section{Three different acceptance conditions}
        \label{sec:different_accepting}
        Here we present three different acceptance conditions for membrane systems with active membranes and show what complexity class they characterise.
        We define each acceptance condition; define a graph reachability problem that models the computation of such a system; then prove both upper and lower bounds on the computational power of the system.
        Each of Definitions~\ref{def:general_recogniser_membranes}, \ref{def:recogniser_membranes}, \ref{def:strict_recogniser_membranes}, is a more concrete replacement for Definition~\ref{def:vague_recogniser}.
        Most results in this section use reductions to and from reachability problems on membrane dependency graphs.
        Solving these reachability problems is equivalent to simulating such a membrane system since we translate (via $\AC^0$ reductions) from a membrane system to a corresponding reachability problem, and vice-versa.

        \subsection{General recogniser systems characterise $\NL$}
            \label{ssec:general_rec_sys}
            In previous works~\cite{MW2008c,MW2007p} we used a definition of recogniser membrane systems that is more general than is typical of other work in the area (i.e.~Section~\ref{ssec:standard_rec_sys}). 
            In this more general definition it is possible for the membrane system to output both \yes and \no symbols.
            However, when the first of these symbols is produced we call it the accepting/rejecting step of the computation.
            (Note that it is forbidden for both \yes and \no to be produced in the same timestep.)
            We now define this acceptance condition and then go on to show that $\ACPMCwodissSemi$ systems with this acceptance condition characterise $\NL$.

            \begin{defn}
                \label{def:general_recogniser_membranes}
                A {\em general recognizer membrane system}, $\Pi$, is a membrane system with external output (that is, the results of halting computations are encoded in the environment) such that:
                \begin{enumerate}
                    \item the working alphabet contains two distinguished elements \yes and \no;
                    \item if $C$ is a computation of the system, (i) then a \yes or \no object is released into the environment, (ii) but not in the same timestep.
                        If $\yes$ is released before $\no$ then the computation is accepting, otherwise the computation is rejecting.
                \end{enumerate}
            \end{defn}


            \begin{figure}[h]
                \begin{center}
                    \includegraphics{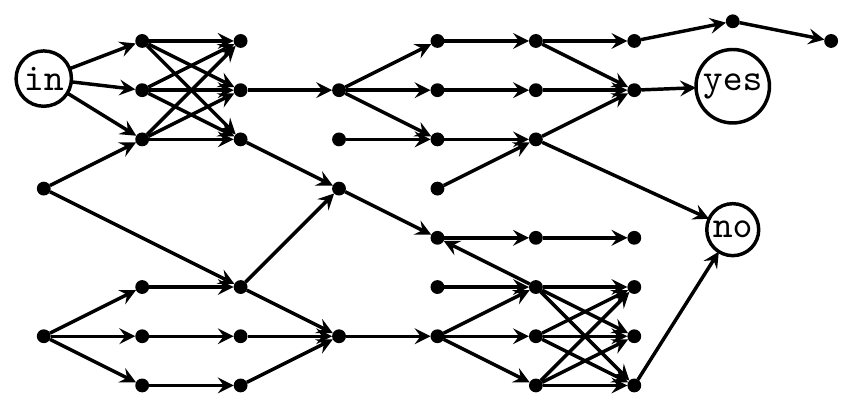}
                \end{center}
                \caption{An example dependency graph $\DG$ for some unspecified {\em general recogniser membrane system} (Definition~\ref{def:general_recogniser_membranes}).
                Note that this represents a rejecting computation since the minimum directed path from $\inputnode$ to $\no$ is of length 6, while the minimum directed path from $\inputnode$ to $\yes$ is of length~7.
                }
                \label{fig:DG_of_general_semi_recogniser}
            \end{figure}

            We now define the reachability problem for $\ACPMCwodissSemi$ systems whose acceptance conditions are as in Definition~\ref{def:general_recogniser_membranes}.
            Solving this problem is equivalent (via a reduction) to simulating such a system. 

            \begin{prob}[\GENERALREACH]
                \mbox{}\\
                \textbf{Instance:} A dependency graph $\DG = \left(\DGv, \DGe, \inputnode, \yes, \no \right)$ where $\left\{ \inputnode, \yes, \no \right\} \subseteq \DGv$, representing the rules of a general recogniser membrane system $\Pi$ as defined in Definition~\ref{def:general_recogniser_membranes}.\\
                \textbf{Problem:} Is the shortest directed path from $\inputnode$ to $\yes$ of length less than the shortest directed path from $\inputnode$ to $\no$?
            \end{prob}

            We also define the problem $\STCON$, the canonical $\NL$-complete problem~\cite{Jon1975p}. 
            This problem is also known as \lang{PATH}, \lang{REACHABILITY}, and \lang{GAP}.
                
            \begin{prob}[\STCON]
                \mbox{}\\
                \textbf{Instance:} A directed acyclic graph $G=(V,E, s, t)$ where $\left\{s, t\right\} \subseteq V$.\\
                \textbf{Problem:} Is there a directed path in $G$ from $s$ to $t$?
            \end{prob}

            We now provide a result which is used to show that \ACPMCwodissSemi systems whose acceptance conditions are as in Definition~\ref{def:general_recogniser_membranes} characterise $\NL$ (this characterisation has been published elsewhere~\cite{MW2008c}, we present a shorter proof here).
            \begin{thm}
                \label{thm:genrec_is_NLc}
                $\GENERALREACH$ is $\NL$-complete
            \end{thm}
            \begin{proof}
                First we show $\STCON \leq_{\AC^0} \GENERALREACH$.
                Given an instance $G=\left(V, E, s, t\right)$ of $\STCON$, we construct a dependency graph $\DG = \left(\DGv, \DGe, \inputnode, \yes, \no \right)$ such that $\DGv = V \cup \left\{\no\right\}$ and $\DGe = E$.
                We replace all instances of $s$ with $\inputnode$, and $t$ with $\yes$, in $\DG$.
                Clearly there is a path from $\inputnode$ to $\yes$ iff there is a path from $s$ to $t$ in $\DG$. 
                We also add a directed path of length $|V|+1$ from $\inputnode$ to $\no$ in $\DG$.
                This ensures that if there is not a path from $s$ to $t$ in $G$, than $\no$ is reached after all other paths have terminated.
                This reduction is computed in $\AC^0$.

                We now prove the correctness of the above reduction.
                Since $\GENERALREACH$ is defined in terms of the general recogniser membrane systems (Definition~\ref{def:general_recogniser_membranes}), we often appeal to Definition~\ref{def:general_recogniser_membranes} in the proof.
                Recall that, via Remark~\ref{rem:dg_to_memsys}, we can translate $\DG$ to a membrane system $\Pi_\DG$ in $\AC^0$.
                \begin{itemize}
                    \item By adding a path of length $|V|+1$ from $\inputnode$ to $\no$ we are guaranteeing that object $\no$ is not produced by the membrane system $\Pi_\DG$ at the same time as {\em any} other object, this satisfies point 2(ii) of Definition~\ref{def:general_recogniser_membranes}.
                    \item If there is a path from $s$ to $t$ in $G$ (and $\yes$ is evolved in $\Pi_{\DG}$) the reduction ensures that a path from $\inputnode$ to $\yes$ exists in $\DG$.
                    Also in either case a path from $\inputnode$ to $\no$ is created by the reduction that ensures the correct output from $\Pi_\DG$.
                    Thus we satisfy point 2(i) of Definition~\ref{def:general_recogniser_membranes}.
                \end{itemize}

                We now show that $\GENERALREACH \in \NL$.
                Let $M$ be a non-deterministic Turing machine with two variables $x$ and $y$.
                Finding the shortest path between two nodes is well known to be computable in $\NL$ via $\leq n$ iterations of a $\STCON$ algorithm.
                Set $x$ to be the shortest path from $\inputnode$ to $\yes$.
                Set $y$ to be the shortest path from $\inputnode$ to $\no$.
                If $x < y$, $M$ accepts, otherwise $M$ rejects.
                Thus $M$ uses a non-deterministic algorithm and two binary counters to solve $\GENERALREACH$ and so the problem is in $\NL$.
            \end{proof}

            \begin{thm}
                $\NL$ is characterised by $\ACPMCwodissSemi$ using the {\em general} acceptance conditions from Definition~\ref{def:general_recogniser_membranes}.
            \end{thm}
            The proof is omitted, but can be obtained by using standard techniques along with Remark~\ref{rem:dg_to_memsys}, Theorem~\ref{thm:genrec_is_NLc}, and Definition~\ref{def:general_recogniser_membranes}.
        
        \subsection{Standard recogniser membrane systems characterise $\NL$}
            \label{ssec:standard_rec_sys}
            In this section we discuss the ``standard'' definition for recogniser membrane systems, i.e.\ the definition that most researchers use when proving results about recogniser membrane systems.
            On a given input, these systems produce either a \yes object or a \no object, but not both.
            Also it is assumed that this occurs in the last timestep of the computation where no other rules are applicable.

            By showing an $\NL$ characterisation for such systems, we are showing that this definition has equal power to the more general definition discussed above in Section~\ref{ssec:general_rec_sys}.
            Furthermore, we have provided a ``compiler,'' via reductions, to translate a system that uses the general definition into a system that uses the standard definition.
            This is significant since the general definition is often easier to program, while it is often easier to prove certain properties (such as correctness) for the standard definition.
            We begin with a definition of standard recogniser membrane systems from Guti{\'e}rrez-Naranjo et al.~\cite{NJNC2006p}.

            \begin{defn}[\cite{NJNC2006p}]
                \label{def:recogniser_membranes}
                A {\em recognizer membrane system}, $\Pi$, is a membrane system with external output (that is, the results of halting computations are encoded in the environment) such that:
                \begin{enumerate}
                    \item the working alphabet contains two distinguished elements \yes and \no;
                    \item all computations halt; and
                    \item if $C$ is a computation of the system, then (i) either object \yes or object \no (but not both) must have been released into the environment, and (ii) only in the last step of the computation.
                    If $\yes$ is released then the computation is accepting, otherwise the computation is rejecting.
                \end{enumerate}
            \end{defn}

            \begin{rem}
                Definition~\ref{def:recogniser_membranes} affects the dependency graph of such systems so that we can define the following subsets of the objects \nmMembraneObjects.\\
                $\nmMembraneObjects_{\yes} = \{ \nmGenericObject\ \vert\ \nmGenericObject \in \nmMembraneObjects \text{ and $o$ eventually evolves } \yes \}$,\\
                $\nmMembraneObjects_{\no} = \{ \nmGenericObject\ \vert\ \nmGenericObject \in \nmMembraneObjects \text{ and $o$ eventually evolves } \no \}$,
                and $\nmMembraneObjects_{\mathrm{other}} = \nmMembraneObjects \backslash (\nmMembraneObjects_{\yes} \cup \nmMembraneObjects_{\no})$.
            \end{rem}
            \begin{lem}
                \label{lem:yes_no_intersect_null}
                $\nmMembraneObjects_{\yes} \cap \nmMembraneObjects_{\no} = \emptyset $.
            \end{lem}
            \begin{proof}
                Assume that object $\nmGenericObject \in \nmMembraneObjects_{\yes} \cap \nmMembraneObjects_{\no}$, this implies that both a \yes and a \no object are produced by the confluent system on a given input which contradicts point 3(i) of Definition~\ref{def:recogniser_membranes}.
            \end{proof}

\begin{figure}[h]
                \begin{center}
                    \includegraphics{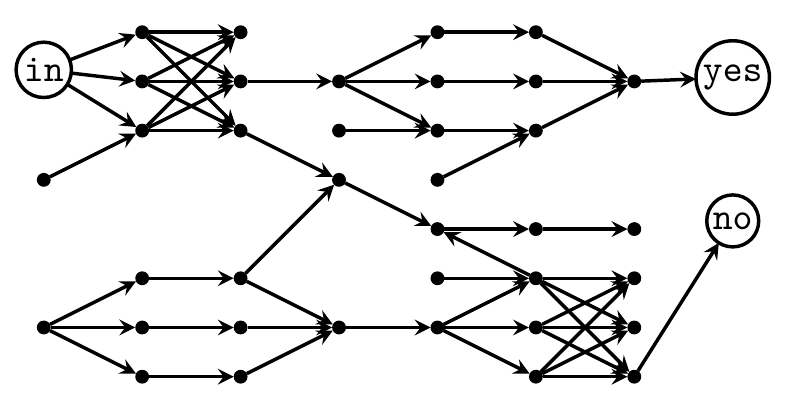}
                \end{center}
                \caption{An example dependency graph $\DG$ for some unspecified standard recogniser membrane system (Definition~\ref{def:recogniser_membranes}).
                        Note that via Lemma~\ref{lem:yes_no_intersect_null} there are no directed paths from $\nmMembraneObjects_{\yes}$ to $\nmMembraneObjects_{\no}$, they are {\em weakly connected}. 
                }
                \label{fig:DG_of_standard_semi_recogniser}
            \end{figure}

            These observations are illustrated in Figure~\ref{fig:DG_of_standard_semi_recogniser}.
            
            %

            We now define the reachability problem for $\ACPMCwodissSemi$ systems whose acceptance conditions are as in Definition~\ref{def:recogniser_membranes}.
            We remind the reader that these systems are confluent via Definition~\ref{def:uniformFam} and Remark~\ref{rem:confluent}.
            \begin{prob}[\STANDARDREACH]
                \mbox{}\\
                \textbf{Instance:} A dependency graph $\DG = \left(\DGv, \DGe, \inputnode, \yes, \no \right)$ where $\{ \inputnode, \yes, \no\} \subseteq \DGv$, representing the rules of a \ACPMCwodissSemi recogniser membrane system $\Pi$ as defined in Definition~\ref{def:recogniser_membranes}.\\
                \textbf{Problem:} Is there a directed path from $\inputnode$ to $\yes$?
            \end{prob}
                
            \noindent We now provide the main result needed to show that standard \ACPMCwodissSemi characterises $\NL$.
            \begin{thm}
                \label{thm:stdrec_is_NLc}
                $\STANDARDREACH$ is $\NL$-complete.
            \end{thm}
            \begin{proof}
                First we show $\STCON \leq_{\AC^0} \STANDARDREACH$. 
                Given an instance $G=\left(V, E, s, t\right)$ of $\STCON$, we construct a dependency graph $\DG = \left(\DGv, \DGe, \inputnode, \yes, \no \right)$ such that $\DGv = V \cup \left\{ \yes, \no \right\}$ and $\DGe = E$.
                We replace $s$ with $\inputnode$ in \DG.
                We add a directed path of $|V|+1$ edges leading from $t$ to $\yes$ to ensure that all other computations have halted before \yes is evolved.
                Clearly there is a path from $\inputnode$ to $\yes$ in $\DG$ iff there is a path from $s$ to $t$ in graph $G$.

                So far, $\DG$ we have shown that $\ACPMCwodissSemi$ recogniser membrane systems, as in Definition~\ref{def:recogniser_membranes}, {\em accept} words in $\STCON$.
                However, the construction does not explicitly say how to reject words that are not in the language, which is a requirement of Definition~\ref{def:recogniser_membranes}.
                We extend the proof as follows.
                Let $\NOTSTCON$ be the complementary problem to $\STCON$, i.e. given an acyclic graph $G'$ is there no directed path from $s'$ to $t'$?
                $\NOTSTCON$ is $\coNL$-complete (via the same reduction that is used to show the $\NL$-completeness of $\STCON$), and so is also $\NL$-complete (since $\NL=\coNL$~\cite{Imm1988p,Sze1987p}).
                Now we define a third $\NL$-complete problem $\STCON\text{--}\NOTSTCON$; the set of graphs with two disjoint components $G, G'$ that are related in the following sense: $s$ eventually yields $t$ in $G$ iff $s'$ does not eventually yield $t'$ in $G'$.
                Now we reduce this graph to a dependency graph $\DG$ in a similar manner as the above reduction.
                That is, we place an edge from $\inputnode$ to $s$ and from $\inputnode$ to $s'$.
                We add a directed path of $|V|+1$ edges leading from $t$ to $\yes$, and another directed path of $|V|+1$ edges leading from $t'$ to $\no$.
                Then the induced membrane system $\Pi_\DG$ correctly decides $\STCON\text{--}\NOTSTCON$ since it answers \yes iff $s$ leads to $t$, otherwise it answers \no. 
                This reduction is computed in $\AC^0$.
                
                We now prove the correctness of the above reduction.
                Recall that, via Remark~\ref{rem:dg_to_memsys}, we translate $\DG$ to a membrane system  $\Pi_\DG$ in $\AC^0$.
                \begin{itemize}
                    \item Since an instance of $\STCON\text{--}\NOTSTCON$ is an acyclic graph we trivially satisfy point 2 of Definition~\ref{def:recogniser_membranes}.
                    \item In the induced membrane system $\Pi_\DG$ the node \inputnode can only lead to one of \yes or \no, but not both, since the embedded $\STCON$ and $\NOTSTCON$ problems are complementary.
                    This satisfies point 3(i) of Definition~\ref{def:recogniser_membranes}.
                    \item $\Pi_\DG$ outputs (either \yes or \no) in the last step because we add $|V|+1$ extra edges from $t$ and $t'$ so that the accepting or rejecting path is the longest in the dependency graph, satisfying point 3(ii) of Definition~\ref{def:recogniser_membranes}.
                \end{itemize}

                Now we show that $\ACPMCwodissSemi$, as in Definition~\ref{def:recogniser_membranes}, can recognise no more than $\NL$ by showing that $\STANDARDREACH \leq_{\AC^0} \STCON$.
                We observe that an instance of $\STANDARDREACH$ is a directed acyclic graph (via point 2 of Definition~\ref{def:recogniser_membranes}).
                Given an instance $\DG = \left(\DGv, \DGe, \inputnode, \yes, \no\right)$ of \STANDARDREACH, we construct $G=\left(V, E, s,t\right)$  such that $V = \DGv$ and $E = \DGe$ and replace all instances of $\inputnode$ with $s$ and $\yes$ with $t$ in $\DG$.
                Clearly there is a path from $s$ to $t$ in $G$ iff there is a path from $\inputnode$ to $\yes$ in the dependency graph $\DG$.
                This reduction is computed in $\AC^0$.
            \end{proof}
            
            \begin{thm}
                $\NL$ is characterised by $\ACPMCwodissSemi$ using the {\em standard} acceptance conditions from Definition~\ref{def:recogniser_membranes}.
            \end{thm}
            The proof is omitted, but can be obtained by using standard techniques along with Remark~\ref{rem:dg_to_memsys}, Theorem~\ref{thm:stdrec_is_NLc}, and Definition~\ref{def:recogniser_membranes}.

        \subsection{Restricted recogniser membrane systems characterise $\L$}
            \label{ssec:strict_rec_sys}
            We now consider a restriction on the standard definition of recogniser membrane systems.
            Above in Section~\ref{ssec:standard_rec_sys}, we forbid an object that eventually yielded a \yes from also yielding a \no (and vice versa).
            Now we further restrict the system and require that {\em all} descendent nodes of $\inputnode$ must eventually yield \yes, or all must eventually yield \no. 
            Notice that this restriction forbids objects that do not contribute to the final answer (accept or reject) and forbids rules of the form $\left[ a \rightarrow \lambda \right]$ where $\lambda$ is the empty word.

            \begin{defn}
                \label{def:strict_recogniser_membranes}
                A {\em restricted recogniser membrane system}, $\Pi$, is a membrane system with external output (that is, the results of halting computations are encoded in the environment) such that:
                \begin{enumerate}
                    \item the working alphabet contains two distinguished elements \yes and \no;
                    \item all computations halt; 
                    \item if $C$ is a computation of the system, then (i) either object \yes or object \no (but not both) must have been released into the environment, and (ii) only in the last step of the computation.
                        If $\yes$ is released then the computation is accepting, otherwise the computation is rejecting.
                    \item each object $\nmGenericObject \in O$ must, via a sequence of zero or more developmental rules, lead to \yes, or else lead to \no, but not both.
                \end{enumerate}
            \end{defn}

            The definition has the following effect on the dependency graph.
            \begin{rem}
                Since every object eventually yields exactly one \yes, or exactly one \no, the graph $\DG$ consists of exactly two disjoint components.
            \end{rem}
            
            \begin{figure}
                \begin{center}
                    \includegraphics{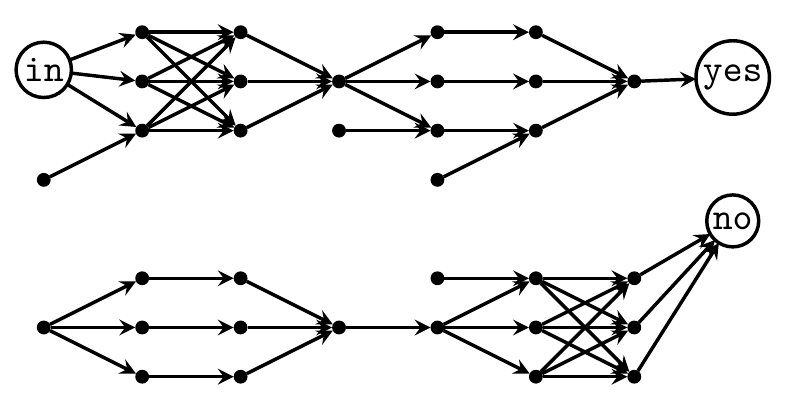}
                \end{center}
                \caption{An example dependency graph $\DG$ for some unspecified restricted recogniser membrane system (Definition~\ref{def:strict_recogniser_membranes}).}
                \label{fig:DG_of_strict_semi_recogniser}
            \end{figure}
                
            We now define a graph reachability problem for $\ACPMCwodissSemi$ systems whose acceptance conditions are as in Definition~\ref{def:strict_recogniser_membranes}.
            \begin{prob}[$\RESTRICTEDREACH$]
                \mbox{}\\
                \textbf{Instance:} A dependency graph $\DG = \left(\DGv, \DGe, \inputnode, \yes, \no \right)$ where $\{ \inputnode, \yes, \no \} \subseteq \DGv$, representing the rules of an \ACPMCwodissSemi recogniser membrane system $\Pi$ as defined in Definition~\ref{def:strict_recogniser_membranes}.\\
                \textbf{Problem:} Is there a directed path from $\inputnode$ to  $\yes$?
            \end{prob}

            \noindent We define the $\L$-complete problem {\sc Directed Forest Accessibility} ($\DFA$)~\cite{CMcK1987p}.
            
            \begin{prob}[$\DFA$~\cite{CMcK1987p}]
                \mbox{}\\
                \textbf{Instance:} An acyclic directed graph $G = (V, E, s, t)$ where $\{s,t \} \subseteq V$ and each node is of out-degree $0$ or $1$.\\
                \textbf{Property:} Is there a directed path from $s$ to $t$?
            \end{prob}

            \begin{thm}
                \label{thm:rstreach_lc}
                $\RESTRICTEDREACH$ is $\L$-complete 
            \end{thm}
            \begin{proof}
                First we show $\DFA \leq_{\AC^0} \RESTRICTEDREACH$.
                Given an instance $G=\left(V, E, s, t\right)$ of $\DFA$, we construct a dependency graph $\DG = \left(\DGv, \DGe, \inputnode, \yes, \no\right)$ such that $\DGv = V \cup \left\{ \no \right\}$ and $\DGe = E \backslash \{(t,v) \vert v \in V\}$.
                We also replace $s$ with $\inputnode$, and add a directed path of length $|V|+1$ from $t$ to $\yes$ in $\DG$.
                Clearly there is a path from $\inputnode$ to $\yes$ in $\DG$ iff there is a path from $s$ to $t$ in graph $G$.
                Note that since we removed the edge (if it exists) leaving $t$, every computation halts (in the induced membrane system $\Pi_{\DG}$) upon evolving $\yes$.
                We also add an edge from all nodes, except \yes, of out-degree 0 to $\no$.
                There is now a path from $\inputnode$ to $\no$ iff there is no path from $s$ to $t$ in $G$ because all paths that do not lead to $\yes$ now lead to $\no$.
                This reduction is computed in $\AC^0$.
                
                We now prove the correctness of the above reduction.
                Recall that, via Remark~\ref{rem:dg_to_memsys}, we translate $\DG$ to a membrane system  $\Pi_\DG$ in $\AC^0$.
                \begin{itemize}
                    \item Since $G$ (as a forest) is acyclic, our reduction ensures $\DG$, and hence any computation of $\Pi_\DG$, is acyclic also, satisfying point 2 of Definition~\ref{def:strict_recogniser_membranes}.
                    \item Our reduction ensures that exactly 2 nodes in $\DG$ have out-degree 0, the (sink) nodes $\yes$ and $\no$, this implies that the only objects that have no applicable rules in $\Pi_\DG$ are $\yes$ and $\no$.
                    This satisfies points 1 and 3(ii) of Definition~\ref{def:strict_recogniser_membranes}. 
                    \item Since every node in $G$ has out-degree 0 or 1, then every node in $\DG$ has out-degree 0 or 1 (and every object in $\Pi_\DG$ has 0 or 1 applicable developmental rules).
                    Combined with the previous point, this implies that all nodes in $\DG$ are on a path to either $\yes$ or $\no$, and that all objects in $\Pi_\DG$ eventually yield either $\yes$ or $\no$, satisfying points 4 and 3(i) of Definition~\ref{def:strict_recogniser_membranes}.
                \end{itemize}

                Now we show $\RESTRICTEDREACH$ is contained in $\L$ by outlining a deterministic logspace Turing machine $M$ that decides $\RESTRICTEDREACH$.
                The input tape of $M$ encodes an instance $\DG = (\DGv, \DGe, \inputnode, \yes, \no)$ of $\RESTRICTEDREACH$.
                Starting with the input node $\inputnode$, $M$ stores this node in a variable called $x$ on its work tape.
                If $x$ is neither $\yes$ nor $\no$ then $M$ searches the set of edges $\DGe$ on its input tape, upon finding an edge $(x,v)$, the machine sets $x$ to be $v$ (overwriting the previous value).
                The computation carries on in this fashion until either $x$ equals $\no$ causing $M$ to reject, or $\yes$, in which case $M$ accepts.

                The algorithm correctly decides $\RESTRICTEDREACH$ because each node in the data-structure has out-degree 0 or 1 and we simply trace along a path until we reach a sink.
                If the sink is \yes, we accept, otherwise we reject.
                Since only one node is stored on $M$'s work tape at any time, $M$ uses $O(\log n)$ space (where $n$ is the input length).
                Thus $\RESTRICTEDREACH \in \L$.
            \end{proof}

            \begin{thm}
                $\L$ is characterised by \ACPMCwodissSemi using the {\em restricted} acceptance conditions from Definition~\ref{def:strict_recogniser_membranes}.
            \end{thm}
            The proof is omitted, but can be obtained by using standard techniques along with Remark~\ref{rem:dg_to_memsys}, Theorem~\ref{thm:rstreach_lc}, and Definition~\ref{def:strict_recogniser_membranes}.

    \section{Conclusions}
        In this paper we have shown how the acceptance conditions of membrane systems affect the computational complexity of the system.
        We have presented an analysis of three different acceptance conditions and proved that they each characterise one of two logspace complexity classes, $\NL$ or $\L$.

        In our previous work~\cite{MW2008c} we used Definition~\ref{def:general_recogniser_membranes} as our acceptance condition.
        Systems using this definition are relatively easy to program (construct a membrane system to solve a problem) because one is not concerned with ensuring the system halts or that only $\yes$ or only $\no$ is output.
        However Definition~\ref{def:recogniser_membranes} is the more common definition that is used when discussing active membrane systems as it is easier to prove correctness for these systems.
        The results in Sections~\ref{ssec:general_rec_sys} and \ref{ssec:standard_rec_sys} reveal that when working with \ACPMCwodissSemi systems, both of Definitions~\ref{def:general_recogniser_membranes} and~\ref{def:recogniser_membranes} characterise $\NL$.
        Our result gives an $\AC^0$ computable compiler to turn a system obeying one definition into a system obeying the other definition.
        This makes the choice of either definition a matter of taste and convenience.

        We also have given the first complexity class defined by membrane systems that characterises $\L$. 
        
        It is interesting to note that the rules of \ACPMCwodissSemi systems allow for the generation of an exponential amount of objects and membranes.
        However these systems decide only those problems that a (non-)deterministic Turing machine uses logarithmic space to decide.

        Here we looked at a number of acceptance conditions for active membrane systems and then characterised the computational complexity classes of the systems.
        However, it is also possible to go in the other direction, that is, to choose a complexity class and then try to engineer an acceptance condition in order to characterise the class.
        This technique may give rise to interesting new characterisations.
        Furthermore, we would hope that it may even be useful to help solve some open questions on the power of certain classes of membrane systems.

        We intend to extend this research to see what effect, if any, acceptance conditions have on the complexity of {\em uniform} active membrane systems.
        The techniques may also prove useful for exploring other classes of membrane systems such as tissue P-systems.
        
    
    \section*{Acknowledgements}
        We would like to thank Mario J. P{\'e}rez-Jim{\'e}nez and Agust{\'i}n Riscos-N{\'u}{\~n}ez for interesting discussion and clarification on the standard definition of recogniser membrane systems.
        We would also like to thank Petr Sos{\'i}k for his comments on an earlier draft of this paper.
        
    \bibliographystyle{eptcs} 

\end{document}